# Direct Fabrication of Atomically Defined Pores in MXenes

Matthew G. Boebinger*,[1] Dundar E. Yilmaz*,[2] Ayana Ghosh,[3] Sudhajit Misra,[1] Tyler S. Mathis,[4] Sergei V. Kalinin,[5] Stephen Jesse,[1] Yury Gogotsi,[4] Adri C. T. van Duin,[2] and Raymond R. Unocic[1]

1. Center for Nanophase Materials Sciences, Oak Ridge National Laboratory, Oak Ridge, TN, 37831 USA

2. Department of Mechanical Engineering, The Pennsylvania State University, University Park, PA 16802, USA

3. Computational Sciences and Engineering Division, Oak Ridge National Laboratory, Oak Ridge, TN 37831, USA

4. A.J. Drexel Nanomaterials Institute and Department of Materials Science and Engineering, Drexel University, Philadelphia, PA, 19104 USA

5. Department of Materials Science and Engineering, University of Tennessee, Knoxville, TN, 37996 USA

## Abstract

Controlled fabrication of nanopores in atomically thin two-dimensional material offers the means to create robust membranes needed for ion transport, nanofiltration, and DNA sensing. Techniques for creating nanopores have relied upon either plasma etching or direct irradiation using electrons or ions; however, aberration-corrected scanning transmission electron microscopy (STEM) offers the advantage of combining a highly energetic, sub-Å sized electron beam for atomic manipulation along with atomic resolution imaging. Here, we utilize a method for automated nanopore fabrication with real-time atomic visualization to enhance our mechanistic understanding of beam-induced transformations. Additionally, an electron beam simulation technique, Electron-Beam Simulator (E-BeamSim) was developed to observe the atomic movements and interactions resulting from electron beam irradiation. Using the 2D MXene $Ti_3C_2T_x$, we explore the influence of temperature on nanopore fabrication by tracking atomic transformation pathways and find that at room temperature, electron beam irradiation induces random displacement of atoms and results in a pileup of titanium atoms at the nanopore edge. This pileup was confirmed and demonstrated in E-BeamSim simulations around the small, milled area in the MXene monolayer. At elevated temperatures, the surface functional groups on MXene are effectively removed, and the mobility of atoms increases, which results in atomic transformations that lead to the selective removal of atoms layer by layer. Through controllable manufacture using e-beam milling fabrication, the production and then characterization of the fabricated defects can be better understood for future work. This work can lead to the development of defect engineering techniques within functionalized MXene layers.

Introduction

Constructing materials atom-by-atom or modifying an existing structure, to selectively tune or enhance properties, can open new pathways to engineer materials for new functional applications. In the case of two-dimensional (2D) materials, unique properties emerge as the thickness of the material is confined to the size of a single to a few atomic layers.[1-6] While top-down, bottom-up, or exfoliation-based synthesis methods have been extensively used to fabricate 2D materials, further modification of the properties of 2D materials can also be accomplished by modifying the atomic structure through controlled electron or ion beam irradiation to create functional defects and nanopores.[4-13] This approach relies partly upon a process known as knock-on damage, wherein a highly energetic source of electrons (or ions) possessing a sufficient amount of energy, interact with the nucleus of atoms or introduce local excitations within the lattice that can result in movement and displacement of atoms.[14-18] Depending upon the physics of the electron/ion impact events, a variety of defects ranging from single atomic vacancies or atomic defects to nanopores can intentionally be created.[2, 6-11] From a practical perspective, processing using this approach has proven to be beneficial for several technological applications, such as highly selective ion transport membranes,[19, 20] gas molecular filters,[21, 22] solid-state nanopore devices,[23-27] and water desalination membranes.[28-32]

There are a limited number of experimental electron beam (e-beam) and ion beam processing techniques that have the capability of visualizing materials at the atomic scale, let alone modifying the atomic structure through precise and controlled placement of individual atoms or introducing functional atomic-scale defect. Aberration-corrected scanning transmission electron microscopy (STEM) inherently possesses the spatial resolution needed to image materials at the atomic scale and has recently been used to enable atomic fabrication,[3-5] where the focused e-beam can either be manually positioned on individual atoms or raster scanned across the specimen to induce transformations. Examples include manipulation and placement of individual dopant atoms,[2, 5-7, 9, 10, 33] and fabrication of nanopores,[34, 35] nanowires,[36-38] and creation of unique atomic edge structures.[34, 39] For the ultimate case of single-atom manipulation, it has been demonstrated that dopant atoms in graphene can be positioned by knocking out adjacent carbon atoms, thereby allowing individual dopant atoms a place to jump into through direct exchange.[5-8, 11, 12, 40] For e-beam-induced transformations in transition metal dichalcogenides (TMDs), the highly energetic e-beam first displaces the chalcogen atoms due to lower binding energy compared to the heavier transition metal atom, resulting in unique nanopores with edge decorated atomic structures.[14-17, 34, 41] In this study, molecular dynamics (MD) simulations were conducted to make a direct comparison of the simulated atomic deformation with experimental e-beam fabrication. In a traditional MD simulation, one of the variants of the Verlet algorithm is used to discretize the time domain and solve Newton's equations of motion. The discrete time interval, often called the MD time step, is usually set in the order of femtoseconds to optimize the accuracy and efficiency of the time integration. This time step choice prevents the simulation of fast particles such as accelerated electrons or ions. Thus, instead of treating fast electrons or ions as point particles in an MD simulation, one can include the interaction between these fast particles with nuclei in the system during the simulation. Following this approach, the Electron-Beam Simulator (E-BeamSim) was developed, which is software designed to drive reactive MD simulations and implement electron-matter interactions.

In this study, real-time visualization and controlled feedback of the atomic fabrication processes were utilized to understand the fundamental mechanisms of atomic transformations in real time at atomic resolution. To do this an automated, feedback-controlled scanning position system coupled with an aberration corrected STEM was utilized to controllably induce atomic transformations while visualizing and tracking atomic displacement events. As a model system, we focus on the MXene $Ti_3C_2T_x$. In the past decade, MXenes have emerged as a new and important class of 2D materials.[42-45] MXenes are based upon transition metal carbides, carbonitrides and nitrides and their general structural formula is given as $M_{n+1}X_nT_x$, where n is the number of atomic layers, M is an early transition metal, X is carbon or nitrogen, and $T_x$ is the surface terminated functional group chemistry (e.g. hydroxyl, oxygen, or fluorine). With interest in MXenes coming from a variety of research areas due to their advantageous electrical and mechanical properties, further research into how functional defects can be intentionally introduced in MXenes is required. When compared to defects produced in other 2D materials under similar conditions, $Ti_3C_2T_x$ possesses a complex structure that provides a more robust monolayer. Previous work conducted by Sang et al. examined atomic transformations of $Ti_3C_2T_x$ experimentally using *in situ* heating STEM as well as simulated the structural changes using DFT calculations on the migration and diffusion barriers and ReaxFF molecular dynamic simulations showing the structural evolution of $Ti_3C_2T_x$ MXene. This study was conducted to observe and study the universal behavior seen where epitaxial TiC layers grow on the surface of the MXene as a result of heat, however, these structural changes were not very controllable locally. In this work, we perform automated *in situ* atomic fabrication experiments in monolayer $Ti_3C_2T_x$ to study the atomic transformations that lead to the creation of defects and nanopores as a function of temperature. At room temperature the milling results in the pileup of the Ti atoms around the edges of the nanopores resulting in disordered edges of the nanopore, which was confirmed using the E-BeamSim simulations. However, when heated to 700 °C it was found that, during the milling, the $Ti_3C_2T_x$ MXene goes through metastable phase transformation before the formation of a nanopore. This metastable phase was found to be controllable and stable using the feedback-controlled beam system.

## Results and Discussion

### *Experimental Electron Beam Fabrication Process*

An illustration of the atomic manipulation process can be seen in Figure 1. While previous studies have used the capabilities of this external feedback-controlled scan positioning system for general image acquisition and directed crystallization of materials,[40, 46] in this study we utilized this e-beam control system as an atomic level subtractive manufacturing tool to permit precision control over the raster area coupled with automated feedback to visually track atomic reordering during the milling procedure. This *in situ* visualization can lead to key insights into the atomic disordering processes that are associated with the formation of defects, nanopores, and metastable phases. This system accomplishes this by moving the e-beam in a programmed spiral motion that was predetermined with various set parameters. These include control over the speed of the spiral scan as well as the radius of the spiral, the duration of a single spiral scan, the direction of the spiral, and the maximum number of spiral scans performed. A spiral scan path is used where a single frequency drive signals the scan coils as it was shown, to result in circular scan images with less scan distortion than conventional raster scans.[47, 48] The beam path that these spirals take can be modified, but the default shape is a circular spiral path since this results in the least distortions and eliminates the need for a flyback duration used in traditional STEM rastering. This elimination of the flyback time allows for a more even dosage of the field of view during the milling process. By controlling the size and the speed of the spiral scan, along with the controls provided by the microscope, the probe dwell time and the electron dose over the scan area can be tailored as well.[3, 34, 46-48]. A schematic of the feedback-controlled loop system is shown in Figure 1a. Before experimenting, the desired size and scan number are selected. As the milling occurs the signal from the detectors is sent to the external beam control system and the frame is captured. The average intensity is then plotted, with the dip in intensity indicating the removal of material while plateaus indicating the formation of the intermediate phase. After reaching a predefined lower threshold intensity, a nanopore is formed and the beam moves to the next location with a new set of milling parameters. The insert shows the initial and final steps of one example milling process. The initial $Ti_3C_2T_X$ MXene monolayer structure, which consists of alternating layers of Ti and C can be seen in the isometric and side view schematic of the atomic structure in Fig 1b and the experimental image in Fig 1d. The beam is then scanned in the spiral pattern over the monolayer as the milling process proceeds (Fig 1c) resulting in an ordered removal of Ti layers while at elevated temperatures (Fig 1e), the process of which will be discussed later in more detail.

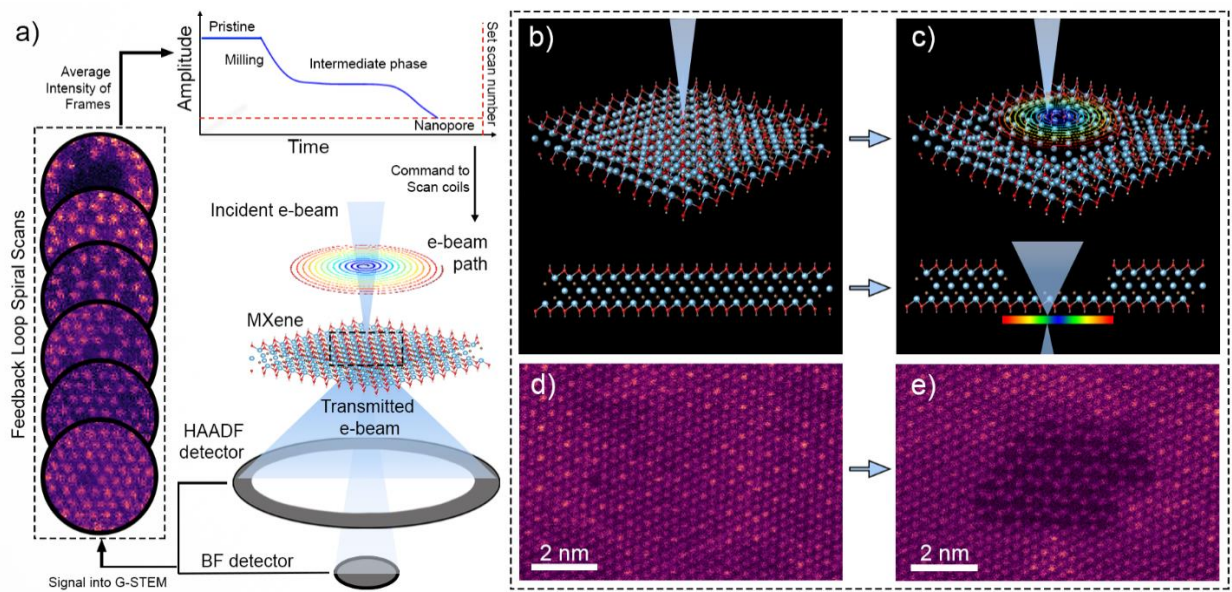

**Figure 1: Illustration of the image-based feedback control method for automated atomic fabrication: (a)** Schematic using the feedback-controlled beam control system that receives input from the detectors to monitor the milling process *in situ* and then, using predefined parameters, automatically controls the experiment by providing commands to scan coils. **(b-c)** Schematic of the model of pristine and milled MXene respectively, with (c) showing the beam path used to mill out the areas in the MXene monolayer. **(d-e)** Experimental images of the pristine MXene and the metastable phase at 700°C, respectively. **(e)** The metastable phase was found to be stable at the elevated temperature up to 25 nm$^2$ in size.

*Phase transformations during milling as a function of temperature*

At room temperature, this fabrication process was performed to make a 3x3 grid of milled nanopores within the 2D MXene monolayer. In Figure 2a the sample was milled while varying the number of maximum cycles, and therefore e-beam irradiation, in the y-direction and the overall size of the milled hole in the x-direction. In this experiment, the nanopore in the upper right corner did not undergo enough scans to fully form a nanopore the same size as seen in future rows. A high-resolution image of a post-drilled nanopore at room temperature in Figure 2b shows the pileup of Ti atoms around the nanopore from the brighter contrast. During e-beam irradiation at room temperature, the C atoms are knocked out and the Ti moves to the edge of the nanopore, leading to this pileup. As previously reported, increasing the temperature results in the removal of different functional groups from the surface of the $Ti_3C_2T_x$ MXene,[49] so upon reaching 700°C, all have theoretically been removed. At this point, it was found that the degradation of the remaining $Ti_3C_2T_x$ monolayer proceeds through a very different mechanism compared to room temperature. This is due to the displaced Ti adatoms being more mobile on the surface of the monolayer[50]. This results in the nanopores formed at 700°C in Fig. 2c having sharper edges with no Ti pileup observed as the more mobile Ti led to the epitaxial growth of further TiC layers on the surface of the monolayer between the nanopores. This is the same phenomenon previously reported by Sang et al. where heating led to the growth of similar homoepitaxial layers[50]; however, by utilizing this e-beam control technique, these milled areas could be controlled. It was also found through the *in situ* visualization of the e-beam milling process that, at 700°C, a metastable intermediate phase formed during the milling process. Additionally, this metastable intermediate phase was found to be stable if the milling process was halted and e-beam irradiation over the area controlled. However, electron dose is not a consistent measure to determine when the metastable phases form, and therefore other measures, such as the image feedback enabled through the external beam control system, were needed. Additional e-beam irradiation through traditional rastering was found to cause the structures to fail and form nanopores but holding at the elevated temperature and cooling down to room temperature did not cause additional changes to the metastable phase's stoichiometry. From *ex situ* analysis, the metastable areas were found to be able to be made up to around 5 nm$^2$ in area reliably and there was found to be no snapback of the structure to the $Ti_3C_2T_x$ phase after the milling process.

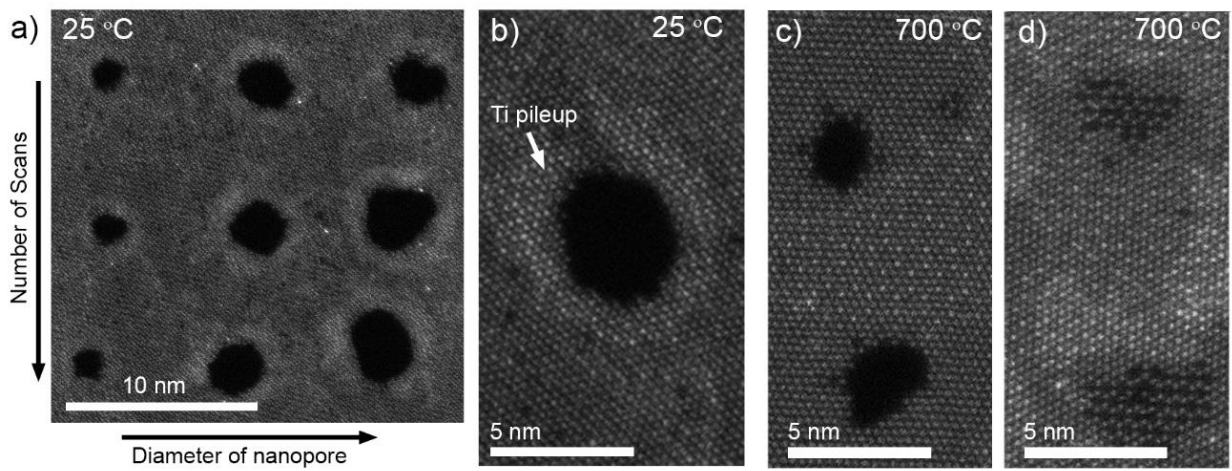

**Figure 2: Nanopore milling of 2D Ti$_3$C$_2$T$_x$ MXene as a function of temperature: (a)** Milling array conducted at room temperature where the size of the nanopores was varied in the x-direction and the number of scans (therefore the electron dose) is varied in the y-direction. **(b)** A high-resolution HAADF-STEM image of a milled nanopore conducted at room temperature where a pileup of Ti atoms is observed. **(c)** A milling experiment conducted at 700°C demonstrates how the Ti grows epitaxial layers on the surface rather than piling up at the edge of the nanopore which remains ordered. **(d)** An example of the metastable phases that were able to be formed using the feedback-controlled beam system.

For further analysis of the *in situ* visualization data, the effect that the elevated temperature and removal of the surface functional groups have on the milling process was compared in Figure 3. At room temperature, with surface functional groups still present, the milling is generally disordered in nature as Ti and C vacancies are formed and filled until a certain concentration of the vacancies is reached and a nanopore forms (Figure 3a). The milling from this point can lead to the formation of larger nanopores as Ti continuously piles up at the edges of the nanopore as discussed above. When directly compared to the milling process at 700°C (Figure 3b) the milling process is much more orderly and generally follows an ordered step-by-step process. The process begins when the Ti$_3$C$_2$T$_x$ structure becomes disordered as Ti and C vacancies in the TiC layer on the opposite side of the incident beam form. The next step in this process involves these V$_{Ti}$ coalescing together into a large, ordered area as the V$_{Ti}$ are all formed on the same layer of Ti. During this coalescing step, the V$_{Ti}$ form along ordered hexagonal crystallographic directions, leaving ordered atomic scale defects with the Ti$_2$C structure. At this point during the milling the experiment can be automatically halted as the intermediate metastable phase has a telltale plateau in the feedback intensity (schematic in Fig. 1a). If halted, the Ti$_2$C metastable phase was found to remain intact with no snapback to the Ti$_3$C$_2$T$_x$ structure, forming a nanowell-like structure within a MXene monolayer similar to that seen previously in multi-layered or stacked 2D materials such as WS$_2$.[51] However, if the milling process proceeds, additional V$_{Ti}$ are formed in the former middle layer of the Ti$_2$C structure resulting in the single layer of Ti structure seen in the middle of the bottom row of Figure 3b. This structure is not stable however and quickly leads to the knockout of the Ti atoms in the final Ti layer. At this point, the nanopore grows rapidly to the furthest extent of the Ti$_2$C metastable phase edges. This step-by-step milling process of the ordered removal of planes one by one can be thought of as similar to a deli-slicer where a full layer seems to need to be removed before the removal of the next. From Figure 3c, we confirmed that this was the process that the milling underwent through the use of multislice STEM simulations. As the experimental images closely match the simulated images of the monolayer $Ti_3C_2T_x$ structure as layers are removed, this clearly shows the removal of the layers of the Ti within the MXene monolayer structure.

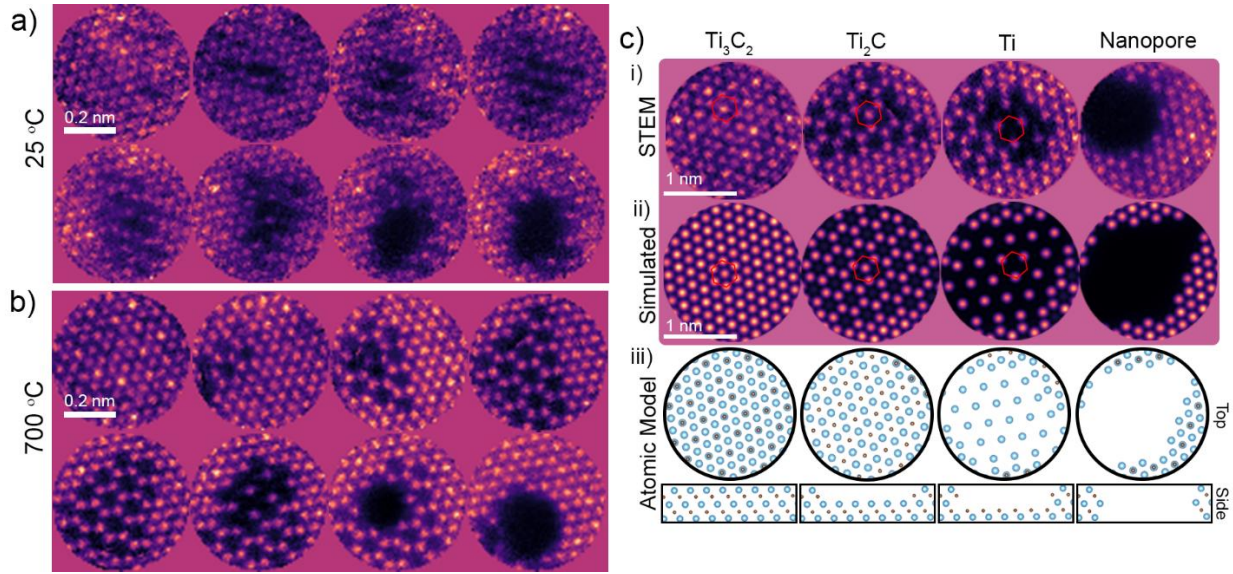

**Figure 3: ADF-STEM images acquired during the spiral scan milling process: (a)** At room temperature the Ti atoms become disordered as the nanopore is formed and expands. **(b)** At 700°C the e-beam material degradation follows a much more ordered process. In these scans, the $V_{Ti}$ form together to form the metastable phases consisting of 2 Ti layers. $V_{Ti}$ then continue to form the 1 Ti layer for a short amount of time before the nanopore fully forms. **(c)** A comparison between the (i) experimental spiral scans, (ii) simulated images, and (iii) atomic structure model showing the progression of the layer-by-layer removal process during milling.

*Barrier energy calculations as a function of temperature*

The difference in the milling process was further explored using density functional theory (DFT) and ab-initio molecular dynamics (AIMD) computations at 0 K and 972 K, respectively. The nudged elastic band (NEB) barrier energy calculations were performed to find the temperature-dependent transition barriers. This was done by extracting the step-by-step structures going from $Ti_3C_2T_x$ to single Ti before nanopore formation from the previous analysis. The STEM-simulated structures were utilized as the initial structures to conduct the computations. First, the optimized structure for $Ti_3C_2$ was obtained from 0 K DFT computations. The intermediate and the final systems of $Ti_2C$ and Ti of the reactions were designed by separating (distance of 20 Å and 10 Å) two layers of TiC, one by one, from the structure of $Ti_3C_2$ and $Ti_2C$, respectively. We performed the NEB computations for the two-steps reaction, first going from $Ti_3C_2$ to $Ti_2C$ followed by $Ti_2C$ to Ti. The intermediate structures were obtained by interpolating each of the initial and final systems for each reaction.

Using the simulated structures as a base, with the exterior edges constrained to act as the structure outside the milling FOV, the structure is relaxed at each known reaction step, being $Ti_3C_2$, $Ti_2C$, and Ti, as shown in Figure 4. The reference energy is that of $Ti_3C_2$ used to evaluate the energy differences. The relative energy from the $Ti_3C_2$ structure to $Ti_2C$ is 4.7 eV/f.u. at lower temperature and 4.9 eV/f.u. at the elevated temperature, with the relative energy from the $Ti_2C$ structure to Ti found to be 4.9 eV/f.u. at the lower temperature and 5.1 eV/f.u. at the elevated temperature. The shape of the NEB in the elevated temperature calculations shows a local minimum at the $Ti_2C$ phase that is not generally seen at lower temperatures. However, there does exist a less pronounced plateau that theoretically shows that the $Ti_2C$ phase is an intermediate phase in the milling process. However these theoretical calculations are performed on non-functionalized surface MXene structures, combined with the fact that the milling experiments do not show this intermediate phase at room temperature, it can be concluded that the presence of the surface functional groups plays a larger part in the formation of this unique structure as the $Ti_2C$ is only seen at elevated temperatures after the surface functional groups are removed.

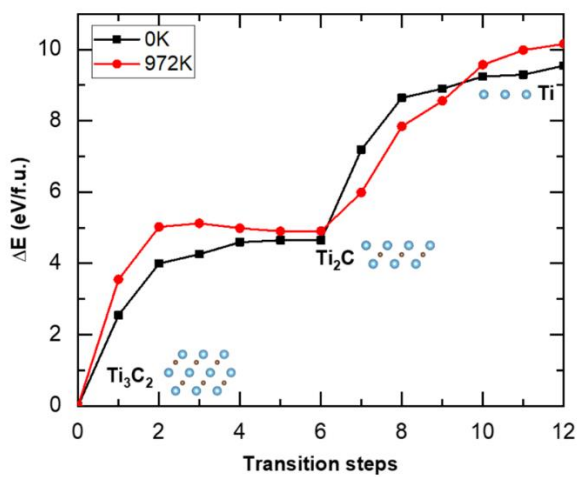

**Figure 4:** Plot of the NEB calculations during the phase transformation from the relaxed $Ti_3C_2$ structure to the intermediate $Ti_2C$ phase and final Ti phase as a function of the temperature with the lower in black and the elevated temperature in red.

*Electron Beam Simulations of $Ti_3C_2T_x$ MXene Monolayers*

While the beam control STEM system can provide atomic resolution imaging of the $Ti_3C_2T_x$ MXene structure, this image remains a two-dimensional projection of a three-dimensional structure. Therefore E-BeamSim Simulations were conducted to demonstrate the 3D behavior of the Ti pileup that the MXene undergoes at room temperature as a result of e-beam irradiation damage. This technique simulated the structure by providing an accurate reactive force field trained with high-fidelity DFT data as described in the Supplementary Data section. In Figure 5, we present snapshots of different perspectives of the MXene monolayer consisting of the $Ti_3C_2T_x$ terminated with O on each surface after 1.2 ns electron beam scan. The perspective (Figure 5a) and top view (Figure 5b) of the initial state of the MXene monolayer with a plot tracking the number of atomic species moved to the surface. After an extended period of e-beam irradiation to the center region and nanopore is shown to form (Figures 5c and 5d). From the perspective view the initial pileup of the Ti along the edge of the nanopore can be observed to be similar to that seen in the experimental millings in Figure 2. After a time, the simulation was stopped and a small stable nanopore was shown (Figures 5e and 5f) to form around the same size as those shown in the left column of Figure 2.

The defect evolution of the monolayer MXene samples was investigated frame-by-frame using E-BeamSim considering different versions of irradiated active area of 1.5 x 1.5 nm. In Figure 5d, the states of the disordering closely match the circular irradiation seen in Figure 3 until the formation of a more stable structured nanopore. From the plots, it can be seen that a much greater degree of the Ti diffused out onto the surface whereas the more mobile C left the system across the surface in both scenarios. While this system is not fully representative of the surface functional groups present on the MXene monolayer, the E-BeamSim does demonstrate the Ti pileup behavior observed at room temperature even at the much smaller nanopore scale.

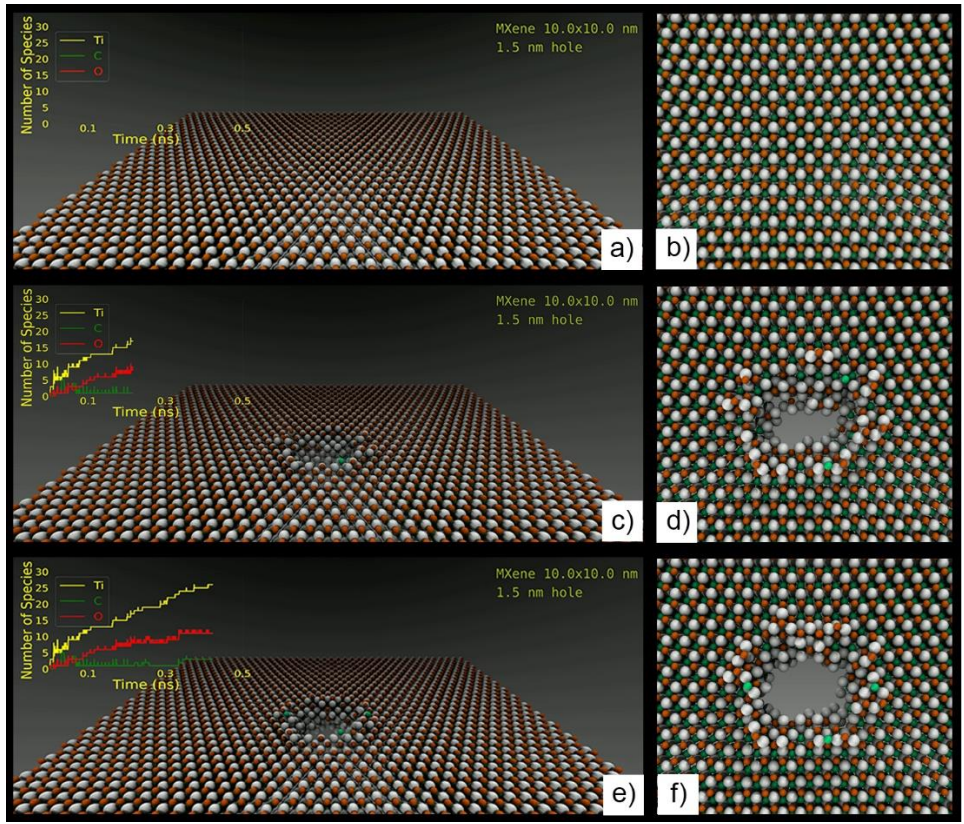

**Figure 5: Different views of Ti$_3$C$_2$T$_x$ monolayer with E-BeamSim: (a-b)** Initial perspective **(a)** and top **(b)** views of a 10x10 nm area Ti$_3$C$_2$T$_x$ MXene structure with an active area of 1.5 nm at the center. Additionally, an inset plot tracks the number and types of atoms diffused to the surface. **(c-d)** Intermediate structure as the nanopore and pileup of the Ti along the edge begins in both the perspective **(c)** and top **(d)** view. **(e-f)** Final defect structure showing the Ti pileup along the surface in both the perspective **(e)** and top **(f)** view. Color code: Ti = gray, C = green, O = red.

Conclusions

This work demonstrated the controllable formation of desired defect structures within MXene that were formed using an external beam control system. By controlling the feedback intensity threshold *in situ*, the beam control system could be used as a tool to fabricate these structures controllably in the 2D material by automatically stopping the drilling process after having only partially milled areas of the MXene. Previous studies have shown the stable formation of nanowell-type structures in multi-layered 2D materials, however through the feedback-controlled e-beam milling system demonstrated here these nanowells can be formed within the MXene monolayer. The complexity of the MXene leads to a more robust structure that allows for the controllable formation of these types of defects. Density Functional Theory and Ab Initio Molecular Dynamics calculations, as well as larger-scale E-BeamSim e-beam simulations, based on ReaxFF simulations coupled to the McKinley Feschbach equation to describe the atomic velocity increase due to electron beam scattering, have corroborated the behavior seen during these milling processes at room temperature. There is a metastable $Ti_2C$ phase possible, thereby indicating the prevalence of the surface functional groups on the nanopore formation. The ordered degradation of $Ti_3C_2T_x$ MXenes at elevated temperatures as a result of e-beam milling is also corroborated by these calculations. This study indicates that, by using the inputs from E-BeamSim simulations, an automated experimental workflow can be achieved that can fabricate a variety of defect structures within 2D materials.

## Methods

### Synthesis of $Ti_3C_2T_x$ MXene

Single layer $Ti_3C_2T_x$ was synthesized as reported previously.[52] Briefly, the $Ti_3AlC_2$ MAX phase was mixed with etchant and stirred for 24 h at 35 °C. The etchant was a 6:3:1 mixture (by volume) of 12 M HCl, DI water, and 50 wt.% HF. The etched $Ti_3C_2T_x$ was washed with DI water via repeated centrifugation and decantation cycles until the supernatant reached pH ~6. The etched multilayer MXene sediment was then dispersed in a 0.5 M solution of LiCl and then stirred for a minimum of 4 h at room temperature. The MXene/LiCl suspension was then washed with DI water via repeated centrifugation and decantation of the supernatant. The $Ti_3C_2T_x$ solutions were centrifuged for 1 h before the MXene supernatants were collected to ensure the final $Ti_3C_2T_x$ solutions were single-flake.

### In situ Scanning Transmission Electron Microscopy and Image Simulations

$Ti_3C_2T_x$ samples were prepared by drop casting $Ti_3C_2T_x$ colloidal solution onto a commercial microelectromechanical system (MEMS) based in situ heating platform from Protochips, Inc. with C or SiN membrane. For the SiN membrane, arrays of holes were created using a $Ga^+$ ion focused ion beam to ensure that freestanding $Ti_3C_2T_x$ flakes do not interact with the substrate during heating. In situ STEM heating experiments were performed using a Nion UltraSTEM, operating at 100 kV with a beam current of 40 pA and equipped with a spherical aberration (Cs) corrector to achieve the 1 Å spatial resolution. A convergence angle of 31 mrad was used, with HAADF detector inner and outer collection angles of 86 mrad and 200 mrad.

### Theory Calculations

DFT and AIMD simulations GGA were performed using the projector augmented plane-wave (PAW) method and PAW-PBE potential [53] as implemented in the Vienna ab initio simulation package (VASP).[54, 55] The structure optimization was performed by relaxing the atoms steadily toward the equilibrium until the Hellman-Feynman forces are less than $10^{-3}$ eV/Å. All geometry optimization computations were carried out with a 400-eV plane-wave cutoff energy with Monkhorst Pack 2*2*2 k-point meshes.[56] AIMD simulations at 972 K were performed using Nose-Hoover thermostat. All 0 K structures of $Ti_3C_2$, $Ti_2C$ and Ti were used to initialize the AIMD simulations and run for 10 ps until the systems reach equilibrium. These structures were utilized to created intermediate images by linear interpolation of the atomic positions for performing NEB computations at 972 K. E-Beam Sim details are provided in the Supplementary Information.

Acknowledgements

STEM experiments were performed at and supported by the Center for Nanophase Materials Sciences at Oak Ridge National Laboratory, which is a Department of Energy (DOE) Office of Science User Facility. Development of the scan control customization was supported by the U.S. DOE, Office of Basic Energy Sciences, Division of Materials Sciences and Engineering. MXene synthesis and development of E-beam Sim was supported by the Fluid Interface, Reactions, Structures and Transport (FIRST) Center, an Energy Frontier Research Center (EFRC) funded by the U.S. DOE, Office of Science, Office of Basic Energy Sciences. Additional support for the development of the E-beam Sim code and its connectivity to ReaxFF was provided by the National Science Foundation (NSF) through the Pennsylvania State University 2D Crystal Consortium–Materials Innovation Platform (2DCC-MIP) under NSF cooperative agreement numbers DMR-1539916 and DMR-2039351. DFT calculations were supported by INTERSECT Initiative as part of the Laboratory Directed Research and Development Program of Oak Ridge National Laboratory, managed by UT-Battelle, LLC, for the U.S. DOE under contract DE-AC05-00OR22725.

References

1. Li, W., X. Qian, and J. Li, *Phase transitions in 2D materials.* Nature Reviews Materials, 2021. **6**(9): p. 829-846.
2. Dyck, O., et al., *Building structures atom by atom via electron beam manipulation.* Small, 2018. **14**(38): p. 1801771-1801779.
3. Jesse, S., et al., *Directing matter: Toward atomic-scale 3D nanofabrication.* ACS Nano, 2016. **10**(6): p. 5600-5618.
4. Kalinin, S.V., A.Y. Borisevich, and S. Jesse, *Fire up the atom forge.* Nature, 2016. **539**: p. 485-487.
5. Dyck, O., et al., *Atom-by-atom fabrication with electron beams.* Nature Reviews Materials, 2019. **4**(7): p. 497-507.
6. Su, C., et al., *Engineering single-atom dynamics with electron irradiation.* Science Advances, 2019. **5**(5): p. eaav2252.
7. Susi, T., J.C. Meyer, and J. Kotakoski, *Manipulating low-dimensional materials down to the level of single atoms with electron irradiation.* Ultramicroscopy, 2017. **180**: p. 163-172.
8. Zagler, G., et al., *Beam-driven dynamics of aluminium dopants in graphene.* 2D Materials, 2022. **9**(3): p. 035009.
9. Susi, T., et al., *Towards atomically precise manipulation of 2D nanostructures in the electron microscope.* 2D Materials, 2017. **4**(042004): p. 1-9.

10. Susi, T., J.C. Meyer, and J. Kotakoski, *Quantifying transmission electron microscopy irradiation effects using two-dimensional materials.* Nature Reviews Physics, 2019. **1**(6): p. 397-405.
11. Tripathi, M., et al., *Electron-beam manipulation of silicon dopants in graphene.* Nano Lett, 2018. **18**(8): p. 5319-5323.
12. Hudak, B.M., et al., *Directed atom-by-atom assembly of dopants in silicon.* ACS Nano, 2018. **12**(6): p. 5873-5879.
13. Qi, Z.J., et al., *Correlating atomic structure and transport in suspended graphene nanoribbons.* Nano Letters, 2014. **14**(8): p. 4238-4244.
14. Zan, R., et al., *Control of radiation damage in $MoS_2$ by graphene encapsulation.* ACS Nano, 2013. **7**(11): p. 10167-10174.
15. Liu, X., et al., *Top–down fabrication of sub-nanometre semiconducting nanoribbons derived from molybdenum disulfide sheets.* Nature Communications, 2013. **4**(1): p. 1776.
16. Kretschmer, S., et al., *Formation of defects in two-dimensional $MoS_2$ in the transmission electron microscope at electron energies below the knock-on threshold: the role of electronic excitations.* Nano letters, 2020. **20**(4): p. 2865-2870.
17. Komsa, H.-P., et al., *Two-dimensional transition metal dichalcogenides under electron irradiation: Defect production and doping.* Physical Review Letters, 2012. **109**(035503): p. 1-5.
18. Yoshimura, A., et al., *Quantum theory of electronic excitation and sputtering by transmission electron microscopy.* Nanoscale, 2023. **15**: p. 1053-1067.
19. Kang, Y., et al., *2D laminar membranes for selective water and ion transport.* Advanced Functional Materials, 2019. **29**(29): p. 1902014.
20. Kim, S., et al., *Extreme ion transport inorganic 2D membranes for nanofluidic applications.* Advanced Materials, 2022: p. 2206354.
21. Zheng, Z., R. Grünker, and X. Feng, *Synthetic two-dimensional materials: A new paradigm of membranes for ultimate separation.* Advanced Materials, 2016. **28**(31): p. 6529-6545.
22. Liu, G., W. Jin, and N. Xu, *Two-dimensional-material membranes: A new family of high-performance separation membranes.* Angewandte Chemie International Edition, 2016. **55**(43): p. 13384-13397.
23. Sun, Y., et al., *Ultrathin two-dimensional inorganic materials: new opportunities for solid state nanochemistry.* Accounts of Chemical Research, 2015. **48**(1): p. 3-12.
24. Schaibley, J.R., et al., *Valleytronics in 2D materials.* Nature Reviews Materials, 2016. **1**(11): p. 16055.
25. Peng, L., et al., *Two-dimensional materials for beyond-lithium-ion batteries.* Advanced Energy Materials, 2016. **6**(11): p. 1600025.
26. Syamsai, R., et al., *Double transition metal MXene ($Ti_xTa_{4-x}C_3$) 2D materials as anodes for Li-ion batteries.* Scientific Reports, 2021. **11**(688): p. 1-12.
27. Mojtabavi, M., et al., *Single-molecule sensing using nanopores in two-dimensional transition metal carbide (MXene) membranes.* ACS Nano, 2019. **13**(3): p. 3042-3053.
28. Foller, T., H. Wang, and R. Joshi, *Rise of 2D materials-based membranes for desalination.* Desalination, 2022. **536**: p. 115851.
29. Meng, B., et al., *Fabrication of surface-charged MXene membrane and its application for water desalination.* Journal of Membrane Science, 2021. **623**: p. 119076.

30. Cao, Z., V. Liu, and A. Barati Farimani, *Why is single-layer MoS$_2$ a more energy efficient membrane for water desalination?* ACS Energy Letters, 2020. **5**(7): p. 2217-2222.
31. Priya, P., et al., *Machine learning assisted screening of two-dimensional materials for water desalination.* ACS Nano, 2022. **16**(2): p. 1929-1939.
32. Raza, A., et al., *Recent advances in membrane-enabled water desalination by 2D frameworks: Graphene and beyond.* Desalination, 2022. **531**: p. 115684.
33. Dyck, O., et al., *Variable voltage electron microscopy: Toward atom-by-atom fabrication in 2D materials.* Ultramicroscopy, 2020. **211**: p. 112949.
34. Boebinger, M.G., et al., *The atomic drill bit: Precision controlled atomic fabrication of 2D materials.* Advanced Materials, 2023: p. 2210116.
35. Yadav, P., Z. Cao, and A. Barati Farimani, *DNA detection with single-layer Ti$_3$C$_2$ MXene nanopore.* ACS Nano, 2021. **15**(3): p. 4861-4869.
36. Lin, J., et al., *Flexible metallic nanowires with self-adaptive contacts to semiconducting transition-metal dichalcogenide monolayers.* Nat Nanotechnol, 2014. **9**(6): p. 436-42.
37. Murugan, P., et al., *Assembling nanowires from Mo−S clusters and effects of iodine doping on electronic structure.* Nano Letters, 2007. **7**(8): p. 2214-2219.
38. Kibsgaard, J., et al., *Atomic-scale structure of Mo$_6$S$_6$ nanowires.* Nano Letters, 2008. **8**(11): p. 3928-3931.
39. Sang, X., et al., *In situ edge engineering in two-dimensional transition metal dichalcogenides.* Nat Commun, 2018. **9**(1): p. 2051.
40. Jesse, S., et al., *Direct atomic fabrication and dopant positioning in Si using electron beams with active real-time image-based feedback.* Nanotechnology, 2018. **29**(25): p. 255303.
41. Yoshimura, A., et al., *First-principles simulation of local response in transition metal dichalcogenides under electron irradiation.* Nanoscale, 2018. **10**(5): p. 2388-2397.
42. Naguib, M., et al., *Two-dimensional nanocrystals produced by exfoliation of Ti$_3$AlC$_2$.* Advanced Materials, 2011. **23**(37): p. 4248-4253.
43. Naguib, M., et al., *Two-dimensional transition metal carbides.* ACS Nano, 2012. **6**(2): p. 1322-1331.
44. Anasori, B., M.R. Lukatskaya, and Y. Gogotsi, *2D metal carbides and nitrides (MXenes) for energy storage.* Nature Reviews Materials, 2017. **2**(2): p. 16098.
45. Gogotsi, Y. and B. Anasori, *The rise of MXenes.* ACS Nano, 2019. **13**(8): p. 8491-8494.
46. Jesse, S., et al., *Atomic-level sculpting of crystalline oxides: Toward bulk nanofabrication with single atomic plane precision.* Small, 2015. **11**(44): p. 5895-5900.
47. Sang, X., et al., *Precision controlled atomic resolution scanning transmission electron microscopy using spiral scan pathways.* Scientific Reports, 2017. **7**(1): p. 43585.
48. Sang, X., et al., *Dynamic scan control in STEM: spiral scans.* Advanced Structural and Chemical Imaging, 2016. **2**(6): p. 1-8.
49. Hart, J.L., et al., *Control of MXenes' electronic properties through termination and intercalation.* Nat Commun, 2019. **10**(1): p. 522.
50. Sang, X., et al., *In situ atomistic insight into the growth mechanisms of single layer 2D transition metal carbides.* Nat Commun, 2018. **9**(1): p. 2266.
51. Chen, J., et al., *Spatially controlled fabrication and mechanisms of atomically thin nanowell patterns in bilayer WS$_2$ using in situ high temperature electron microscopy.* ACS Nano, 2019. **13**(12): p. 14486-14499.

52. Mathis, T.S., et al., *Modified MAX phase synthesis for environmentally stable and highly conductive $Ti_3C_2$ MXene.* ACS Nano, 2021. **15**(4): p. 6420-6429.
53. Kresse, G. and D. Joubert, *From ultrasoft pseudopotentials to the projector augmented-wave method.* Physical Review B, 1999. **59**(3): p. 1758-1775.
54. Kresse, G. and J. Furthmüller, *Efficient iterative schemes for ab initio total-energy calculations using a plane-wave basis set.* Physical Review B, 1996. **54**(16): p. 11169-11186.
55. Kresse, G. and J. Hafner, *Ab initio molecular dynamics for liquid metals.* Physical Review B, 1993. **47**(1): p. 558-561.
56. Monkhorst, H.J. and J.D. Pack, *Special points for Brillouin-zone integrations.* Physical Review B, 1976. **13**(12): p. 5188-5192.
57. Yan, Q., et al., *Displacement cross sections of electron irradiated graphene and carbon nanotubes.* Nuclear Instruments and Methods in Physics Research Section B: Beam Interactions with Materials and Atoms, 2015. **350**: p. 20-25.
58. McKinley, W.A. and H. Feshbach, *The coulomb scattering of relativistic electrons by nuclei.* Physical Review, 1948. **74**(12): p. 1759-1763.

## Supplementary Information

*Electron-Matter Interactions:*

In a STEM, electrons are accelerated with a voltage range 80 keV to 100 keV however the resulting momentum results in the formation of defects in the sample. Classically, elastic and inelastic scattering are the two mechanisms for electron-matter interaction under scanning transition electron microscope. When considering elastic scattering, relativistic effects are taken into account and are calculated with Rutherford scattering cross section model:

$$\sigma_R = \left(\frac{Ze^2}{2m_e c^2}\right)^2 \frac{1-\beta^2}{\beta^4} \csc^4 \frac{\theta}{2}$$

Where $\beta = V_e/c$ is the ratio of speed of electron to the speed of light, $Z$ is the atomic number of the target nucleus, $m_e$ is the mass of the electron, $e$ is the electron charge, and $\theta$ is the electron scattering angle as seen in Figure S1.

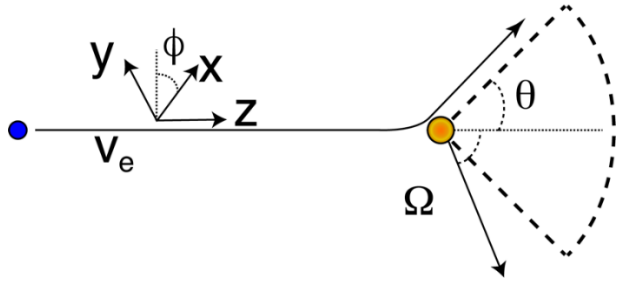

**Figure S1:** Electron scattering from nucleus. $V_e$ is the velocity of the incident electron, θ and Ω are scattering and emission angles of electron and nuclei respectively. φ is the azimuthal angle of the scattering electron.

Rutherford model expanded by Mott into convergent infinite series [57] and, McKinley and Feshbach followed with an approximation to Mott's equation [58]:

Eq. 1.  $\sigma(\theta) = \sigma_R \left[1 - \beta^2 \sin^2 \frac{\theta}{2} + \pi \frac{Ze^2}{\hbar c} \beta \sin \frac{\theta}{2} \left(1 - \sin \frac{\theta}{2}\right)\right]$

The maximum transferred energy to the target atom in elastic scattering of electrons corresponds to the scattering angle $\theta = \pi$:

Eq. 2.  $T_{max} = \frac{2ME(E+2m_e c^2)}{(M+m_e)^2 c^2 + 2ME}$.

Where, $M$ is the mass of target atom and $E$ is the energy of incident electron. The incident electron transfers momentum, thus energy, to the target atom based on the scattering angle. The energy transferred to the nuclei as a function of scattering angle is given as:

$$\text{Eq. 3.} \quad T(\theta) = T_{max} \sin(\tfrac{\theta}{2})$$

Above the threshold required to displace the nuclei from the lattice, the target atom may be removed from its lattice site. On the other hand, if the energy is below this threshold, the transferred momentum will not be sufficient to remove the atom from the lattice and will dissipate as heat. Only electrons that are scattered with an angle greater than the minimum scattering angle will be considered in our simulation scheme. This minimum angle is written as:

$$\text{Eq. 4.} \quad \theta_{min}[u] = 2 \arcsin\left(\sqrt{\tfrac{T_{min}[u]}{T_{max}}}\right)$$

Here $T_{min}[u]$ is the minimal transferred energy to atom $u$. These energies are selected as 80% of the value of the previously reported knock-out potentials. Based on these minimum scattering angles, total cross section of scattering electrons from atom $u$ can be integrated as:

$$\text{Eq. 5.} \quad \sigma_{tot}[u] = 2\pi \int_{\theta_{min}}^{\pi} \sigma(\theta) \sin\theta \, d\theta.$$

The scattering angle $\theta$ of electron is chosen within the interval $\theta \in [\theta_{min}[u_i], \pi]$ according to the probability distribution:

$$\text{Eq. 6.} \quad f(\theta) = \tfrac{2\pi\sigma(\theta)\sin\theta}{\sigma_{tot}[u_i]}.$$

Finally, the emission angle of the target atom can be written as:

$$\text{Eq. 7.} \quad \Omega = \tfrac{\pi - \theta}{2}.$$

The azimuthal angle $\phi$ of the scattering electron selected in the interval $\phi \in [0, 2\pi]$ with a uniform distribution. After selecting the scattering and azimuthal angles of the scattering angle, the momentum transferred to the target atom can be written as:

$$\text{Eq. 8.} \quad \Delta P = \sqrt{2MT(\theta)}\ .$$

The run cycle for the E-BeamSim simulation is run through a python script which drives Lammps/ReaxFF as a MD engine. Simulation starts through initialization of the system through creating a Lammps object and a mesh grid on the sample. Then the mesh further partitioned into three regions: a border of fixed atoms just to avoid moving the sample, a heat-sink region, and then the active region in the center. Atoms in the heat sink region are governed by a thermostat which mimics the dissipation of heat applied by the e-beam as previously described. Finally, at the center is the active region in which e-beam interacts with the sample. In the simulation the electron beam starts moving along the mesh in the active region and at each iteration the E-BeamSim picks a random atom from the current mesh point and then solves the McKinley

Feschbach equation to calculate scattering angle of the electron and velocity added to the target atom.

Next the E-BeamSim simulation adds this velocity to the target atom and returns to the MD cycle. Lammps then relaxes the system for 10 ps before continuing the process for the next mesh point. During the simulations the trajectories of atoms are recorded and then later, images of every frame are rendered and converted into an animation. Additionally the number of atoms sputtered to the surface are tracked and analyzed within the simulation. In this experiment the at 10 x 10 nm area of $Ti_3C_2$ MXene monolayer with surfaces terminated with O was simulated with different size and shape of active areas.

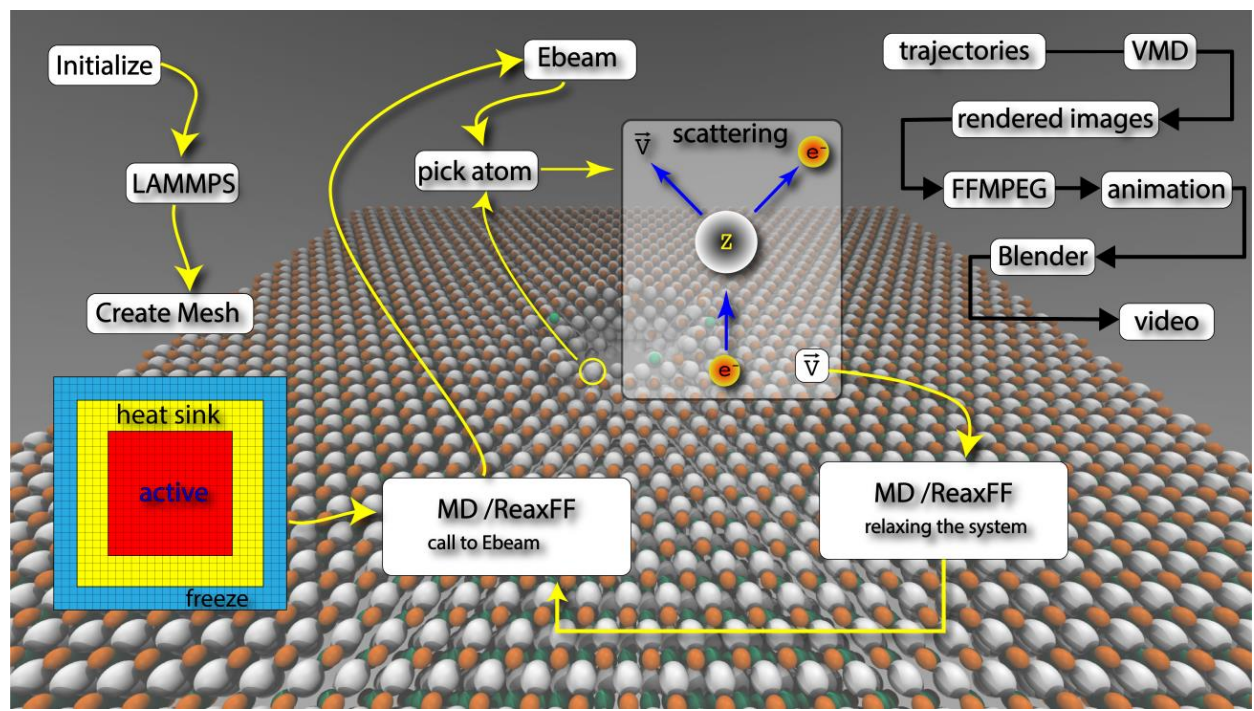

**Figure S2:** The workflow of the E-BeamSim simulation for the $Ti_3C_2$ MXene structure with O terminated surfaces. The structure is first initialized and then heat is applied to the small centralized area and ReaxFF is run on this area to simulate the electron beam interacting with the active area.